\documentclass[%
reprint, superscriptaddress,
amsmath,amssymb, aps, pra,
floatfix, ]{revtex4-1}
\usepackage{nicefrac}
\usepackage{graphicx}
\usepackage{dcolumn}
\usepackage{bm}
\usepackage{todonotes}
\usepackage[protrusion=true,expansion=true,tracking=true]{microtype}

\usepackage{hyperref}
\usepackage[capitalize]{cleveref}

\crefname{section}{Sec.}{Sec.}

\DeclareMathOperator{\dd}{d}
\DeclareMathOperator{\eul}{e}
\newcommand{\imag}{\ensuremath{\text{\textsl{i}}}}

\newcommand{\VC}{\hat{\vec{c}}}       
\newcommand{\VCh}{\VC^\dagger}
\newcommand{\C}{\hat c}       
\newcommand{\hamil}{\hat H}
\newcommand{\Ch}{\C^\dagger}
\renewcommand{\vec}[1]{{\bm{#1}}}
\newcommand{\N}{\hat n}

\newcommand{\bra}[1]{\left\langle #1\right|}
\newcommand{\ket}[1]{\left| #1\right\rangle}
\newcommand{\braket}[2]{\ensuremath{\left\langle #1\middle|#2 \right\rangle}}
\newcommand{\expect}[1]{\ensuremath{\langle #1 \rangle}}
\newcommand{\hc}[0]{\ensuremath{\mathrm{H.c.}}}
\newcommand{\mycaption}[2]{\caption[#1]{\emph{#1} #2}}

\begin{document}
\title{Quantum phases and  topological properties of  interacting fermions in one-dimensional  superlattices}
\author{L.~Stenzel}
\affiliation{Department of Physics,
    Arnold Sommerfeld Center for Theoretical Physics (ASC),
    Ludwig-Maximilians-Universit\"{a}t M\"{u}nchen,
    D-80333 M\"{u}nchen, Germany.}
\affiliation{Munich Center for Quantum Science and Technology (MCQST), Schellingstr. 4, D-80799 M\"unchen, Germany}
\author{A.~L.~C.~Hayward}
\affiliation{Institute for Theoretical Physics, Georg-August-Universit\"at G\"ottingen,
Friedrich-Hund-Platz 1, 37077 G\"ottingen, Germany}
\author{C.~Hubig}
\affiliation{Max-Planck-Institute for Quantum Optics, Hans-Kopfermann-Stra\ss{}e 1, 85748
Garching, Germany}
\author{U.~Schollw\"ock}
\affiliation{Department of Physics,
    Arnold Sommerfeld Center for Theoretical Physics (ASC),
    Ludwig-Maximilians-Universit\"{a}t M\"{u}nchen,
    D-80333 M\"{u}nchen, Germany.}
\affiliation{Munich Center for Quantum Science and Technology (MCQST), Schellingstr. 4, D-80799 M\"unchen, Germany}
\author{F.~Heidrich-Meisner}
\email{heidrich-meisner@uni-goettingen.de}
\affiliation{Institute for Theoretical Physics, Georg-August-Universit\"at G\"ottingen,
Friedrich-Hund-Platz 1, 37077 G\"ottingen, Germany}

\date{March 14, 2019}

\begin{abstract}
The realization of artificial gauge fields in ultracold atomic gases has opened up a path
towards experimental studies of topological insulators and, as an ultimate goal, topological quantum matter in many-body systems.
As an alternative to the direct implementation of two-dimensional lattice Hamiltonians that host the quantum Hall effect and its variants, topological charge-pumping experiments provide an additional avenue towards studying many-body systems.
Here, we consider an interacting two-component gas of fermions realizing a family of one-dimensional superlattice Hamiltonians with onsite interactions and a unit cell of three sites, whose groundstates would be visited in an appropriately defined charge pump.
First, we investigate the grandcanonical quantum phase diagram of individual Hamiltonians, focusing on insulating phases. 
For a certain commensurate filling, there is a sequence of phase transitions from a band insulator to other insulating phases (related to the physics of ionic Hubbard models) for some members of the manifold of Hamiltonians. 
Second, we compute the Chern numbers for the whole manifold in a many-body formulation and show that, related to the aforementioned quantum phase transitions, a topological transition results in a change of the value and sign of the Chern number. 
We  provide both an intuitive and conceptual explanation and argue that these properties could be observed in quantum-gas  experiments. 
\end{abstract}

\maketitle

\section{Introduction}

\label{sec:intro} One of the most prominent examples of topological effects in condensed
matter physics is the quantum Hall effect \cite{vonKlitzing2017}. 
On a lattice, the physics of the quantum Hall effect can be described using a
two-dimensional (2D) square-lattice model pierced by a homogeneous magnetic field, the Harper-Hofstadter
model \cite{Harper1955, Hofstadter1976}. 
Using linear response theory \cite{Kubo1957}, one
can show that the Hall conductivity is quantized \cite{Thouless1982} due to the
{topology} of the bandstructure.

A rather new experimental approach for the  implementation of  topological lattice models is given by ultracold atomic
gases in optical lattices \cite{Galitski2013,Goldman2014, Cooper2018, Aidelsburger2018} in the presence of artificial gauge fields. 
There are different methods to emulate artificial gauge fields, such as lattice shaking \cite{Struck2012,Jotzu2014, Flaeschner2016}, synthetic lattice dimensions \cite{Stuhl2015,Mancini2015, An2017} or laser-assisted tunneling \cite{Aidelsburger2013,Miyake2013}.
 Using the last one, the 
 Harper-Hofstadter model has been realized with ultracold bosons \cite{Aidelsburger2013,Miyake2013,Aidelsburger2015,Tai2017}. 

Accessing the regime of strong interactions while at the same time staying sufficiently close to the groundstate remains a significant experimental challenge. 
This is partly due to the periodic driving used to emulate gauge
fields \cite{Eckardt2017} which leads to heating in generic many-body systems \cite{DAlessio2014,Lazarides2014}.
Significant experimental efforts are geared towards minimizing both systematic and technical sources of heating in multi-band Floquet systems \cite{Reitter2017, desbuquois2017controlling,Messer2018}.

\begin{figure}[tb] 
\centering \includegraphics[width=\columnwidth, clip]{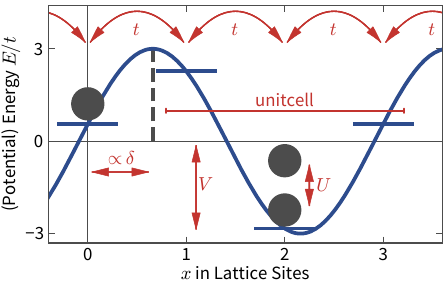}
  \mycaption{(Color online) Schematic representation of the superlattice model.}
  {Fermions hop between adjacent sites with rate $t$ and the onsite potential strength varies cosinusoidally with amplitude $V=3t$. The superlattice potential is invariant under translations by $q=3$ sites and shifted with the phase $\delta$. If there are two particles on one lattice site, energy is increased by $U$.
}
\label{fig:sketch} 
\end{figure}

The aforementioned examples aim at direct implementations of  two-dimensional bandstructures with topological properties.
There is, however, another manifestation of quantum Hall physics: 
By choosing the Landau gauge, the noninteracting 2D Harper-Hofstadter model maps to a family of
uncoupled one-dimensional (1D)  Hamiltonians \cite{Hofstadter1976} that are parameterized by 
a phase $\delta$ (see \cref{fig:sketch} for a sketch of the model). 
Such systems are readily available in several quantum-gas groups \cite{Schreiber2015,Roati2008,Lohse2016,Nakajima2016}.
By adiabatically and periodically changing a set of Hamiltonian parameters, one obtains a Thouless
charge pump \cite{Thouless1983}, in which a quantized amount of charge is transported during each pump cycle.
Such charge pumps have been studied in ultracold atoms by using superlattices realized by two standing-wave laser potentials, 
whose relative phase is varied slowly in order to drive the pump \cite{Lohse2016,Nakajima2016,Schweizer2016} (see \cite{Lohse2018, Zilberberg2018} 
for higher-dimensional versions).

These experiments with (commensurate) superlattices, as well as the theoretical interest in charge pumps,  triggered theoretical investigations of  1D superlattice Hamiltonians and the many-body physics of fermions and bosons in these systems (see, e.g., \cite{Xu2013, Hu2017, Nakagawa2018, Hayward2018}).
The starting point is often the model used in this paper (see the sketch in \cref{fig:sketch}), which  contains a superlattice potential and corresponds exactly to the noninteracting 2D Harper-Hofstadter model. 
Another  paradigmatic model for topological charge pumps is the Rice-Mele model \cite{Rice1982}, which also features spatially dependent hopping amplitudes.

The topological quantization of charge transport in Thouless pumps requires that the system remains in its groundstate as the Hamiltonian parameters are varied. This implies that the many-body groundstate must remained gapped, such that an adiabatic limit is well-defined.
Therefore, there has been great interest in finding insulating quantum phases for both bosonic \cite{Zhu2013,Deng2014,Li2015,Kuno2017,Nakagawa2018} and fermionic \cite{Zhu2013,Hu2017} systems.

Furthermore, one can also establish an analogy to the spin-Quantum Hall effect by studying families of such 1D Hamiltonians with a spin-dependent optical potential \cite{Zhu2013,Hu2017}. 
This requires that one works with two-component gases. 
Recently, anyons have also been studied: 
A variation of  the {statistical} angle, i.e., exchange statistics, can also drive transitions between phases with different topological properties~\cite{Zuo2018}.

In a recent study involving some of us \cite{Hayward2018}, we investigated  a topological charge pump in the interacting bosonic Rice-Mele model.
In that case, interactions of a finite strength are necessary to establish an insulating phase to begin with \cite{Grusdt2013,Rousseau2006,Hayward2018}.
In the case of spinful fermions, starting from the noninteracting case, either one has a band insulator (BI) initially  or works at a half-integer filling (odd number of fermions per unit cell), where in 1D, usually, any arbitrarily small onsite interaction leads to a Mott-insulating state.

In this paper, we study the effect of interactions on a three-band Fermi-Hubbard-Harper superlattice model in a one-dimensional system of spin-1/2 fermions (see \cref{fig:sketch}).  
We discuss the fate of band-insulating phases (BI) and the emergence of Mott insulators (MI) at various fillings and employ density-matrix-renormalization group (DMRG) techniques~\cite{White1992,Schollwoeck2005,Schollwoeck2011} to compute and characterize the grandcanonical phase diagram. 
 
The Hubbard model in the presence of onsite potentials has been studied previously in the context of the ionic Hubbard model \cite{egami1993, fabrizio1999, Takada2001, yamamoto2001, kampf2003, Lou2003, Aligia2004, batista2004, Manmana2004, otsuka2005, lombardo2006, kancharla2006, Torio2006, paris2007, byczuk2009,Tincani2009, garg2014, murcia2016, Loida2017}. 
Most of this attention has been to the two-site periodic potential \cite{egami1993, fabrizio1999,Takada2001,kampf2003,Lou2003,Aligia2004, batista2004, Manmana2004, otsuka2005, lombardo2006, kancharla2006, paris2007, byczuk2009, Tincani2009, garg2014, Loida2017}, although some relevant extensions to the three-site model have also been studied \cite{Torio2006, yamamoto2001, murcia2016}, which apply to our model at certain values of the superlattice phase. 
In the two-site version, it was found that at half filling, the system undergoes a sequence of transitions from a band-insulating state to a correlated insulator (CI) with increasing $U$, with an intermediate spontaneously dimerized insulating (SDI) phase which breaks the 
lattice-inversion symmetry \cite{egami1993, fabrizio1999,Takada2001,kampf2003,Lou2003, Aligia2004, batista2004, Manmana2004, otsuka2005, lombardo2006, kancharla2006, paris2007, byczuk2009, Tincani2009, garg2014, Loida2017}. 

We then use our knowledge of the phase diagram to study the topological properties of various families of adiabatically connected 1D Hamiltonians, parameterized by $\delta$ and a twist angle $\theta$ (introduced via twisted boundary conditions).
We classify these families by a many-body Chern number, an integer-quantized topological invariant.  
In the limit of large systems, the Chern number corresponds to the quantized charge transport in a Thouless charge pump \cite{Thouless1982}, which could readily be carried out in an experiment.

\begin{figure}[t!] \centering \includegraphics[width=\columnwidth, clip]{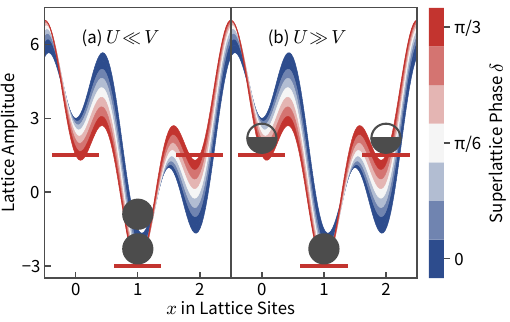}
  \mycaption{(Color online) Schematic representation of band insulator and
    correlated-insulator phases in one superlattice cell.}{These states occur for
    $\alpha=1/3$ and superlattice phase $\delta=\pi/3$ at density $\rho=2/3$. For this
    configuration, the two `upper' lattice sites are degenerate. (a) For
    a weak repulsion $U\ll V$, both particles are localized at the lowest potential
  site. (b) Double occupation is suppressed for strong interactions, $U\gg V$.
}
\label{fig:realspace_model}
\end{figure}

We  find that the presence of quantum phase transitions in our model leads to  interaction-induced changes in the Chern number.
Along with numerical evidence, we provide an intuitive explanation for these topological transitions based on the atomic limit and properties of the bandstructure. An essential aspect is sketched in Fig.~\ref{fig:realspace_model}: depending on whether $U\ll V$ or $U \gg V$,
the lowest site in the unit cell is doubly or singly occupied, respectively. 
This behavior survives away from the atomic limit, in the sense that one can think of the $U\ll V$  case as a (doubly) filled lowest band, while in the $U\gg V$ case, the two lowest bands are effectively filled with only one component.   
These situations translate into different total Chern numbers on finite systems.

Furthermore, we show that the topological structure can be understood in terms of degeneracies associated with the transition between symmetry-protected topological phases driven by the Hubbard interaction. 
We note that the interpretation as a topological charge pump may not be justified in the thermodynamic limit since there exist parameter regions without a  global many-body gap, as required for the adiabatic charge pumping. 
These regions occur for special values of the superlattice phase $\delta$, as will be discussed later, and exhibit vanishing spin gaps resulting in  globally gapless states.
We remind the reader that in ultracold quantum-gas experiments, we are dealing with finite particle numbers that are comparable to what we reach in our numerical simulations.

Our results agree with previous literature wherever we overlap~\cite{Xu2013, Hu2017, Nakagawa2018}. 
Furthermore, a similar transition from a band to a strongly correlated  insulator was observed in the spin-imbalanced case in the same model in \cite{Hu2017}, in this case leading to a change in the spin Chern number.

The paper is structured as follows: In \cref{sec:model}, we start by introducing the Fermi-Hubbard-Harper model in detail and discuss symmetries of the model. 
The following \cref{sec:methods} outlines our numerical method, the DMRG algorithm, and we describe observables studied in this paper. 
In \cref{sec:groundstates}, we present a grandcanonical phase diagram for the Fermi-Hubbard-Harper model, and we discuss physical properties of the insulating phases for {single} Hamiltonians $\hamil(\delta,\theta)$ of the family. 
Then, in \cref{sec:res_topo_props}, we discuss topological properties of the family of groundstates for $\{\hamil(\delta, \theta)\}_{\mathbb{T}^2}$. 
We conclude this exposition with a summary and an outlook contained in \cref{sec:sum}.

\section{Fermi-Hubbard-Harper model}
\label{sec:model} We consider a one-dimensional tight-binding chain with
spin-$1/2$ fermions.  
The model \cref{eq:fhh} is expressed in terms of real-space fermionic annihilation (creation) operators $\C^{(\dagger)}_{j,\sigma}$ and particle-number operators
$\N_j=\N_{j,\uparrow} + \N_{j,\downarrow}$, acting on site $j$ on spin component $\sigma$,
\begin{multline}
\hat H(\delta) =-t\left[\sum_{j=0,\sigma}^{L-1} \Ch_{j,\sigma} \C_{j+1,\sigma} +\hc\right]\\ + V\sum_j \cos(2\pi 
\alpha j+\delta) \N_j + U \sum_j \hat{n}_{j, \uparrow}\hat{n}_{j, \downarrow}\,,
\label{eq:fhh}
\end{multline}
where $L$ is the number of sites. Here, $t$ is the nearest-neighbor tunneling strength, and $U$ is the onsite Hubbard interaction. 
Additionally, there is a commensurate superlattice potential with amplitude $V$, and wavenumber $\alpha\equiv p/q\in\mathbb{Q},\, p$ and $q$ coprime. A schematic representation of the model is shown in \cref{fig:sketch}.

We consider both open boundary conditions, $\C_{L,\sigma}=0$ and twisted boundaries, $\C_{L,\sigma}=\eul^{\imag\theta}\C_{0,\sigma}$
The twist angle $\theta \in [0,2\pi)$ corresponds to a flux piercing of the ring.
Periodic boundary conditions are obtained for $\theta=0$.
We define total electron density, $\rho = N/L$, 
 which is related to the filling factor by a factor of two, owing to the spin degree of freedom. $N= \sum_j \langle \hat n_j \rangle$ is the total number of particles. 
The insulating states of interest appear at  commensurate densities, i.e.,  $\rho = l/q$ for some integer $0 \le l \le 2q$, which corresponds to $l$ total fermions per unit cell. 

Throughout this paper, we are also interested in families of Hamiltonians  parameterized by $\delta,\theta \in [0,2\pi) $ defining a torus.
We will refer to such families as $\{\hamil\left( \delta, \theta \right) \}_{\mathbb{T}^2}$.

\subsection{Correspondence to Harper-Hofstadter model}
The commensurate superlattice can be motivated from the noninteracting, two-dimensional
Harper-Hofstadter model:
By working in the Landau gauge, the system can be Fourier transformed along one axis \cite{Hofstadter1976}.
After transforming into the quasi-momentum basis, the system is separable into a set of one-dimensional lattice models, each parameterized by the quasi-momentum $k_y$.

Therefore, we can interpret it as a family of decoupled chains, where the periodic potential stems from the increased unit cell due to the magnetic flux.
Not taking interactions $\propto U$ into account, \cref{eq:fhh} is thus a {hybrid-space} representation of the Hofstadter model: 
$\delta=a_y k_y$ is the position in $y$-momentum space, where $a_y$ is the lattice spacing and $V=2t_y$ corresponds to the hopping rate along that direction.

Expressing the onsite interaction term $\propto U$ in the original 2D Harper-Hofstadter picture, the interaction would be semi-`local' in hybrid-space. 
That is, the repulsion is onsite in the $x$ direction but infinite-range along the $y$ direction. 
Such interactions are not found in traditional electronic materials, 
however, anisotropic interactions could possibly be implemented using synthetic lattice dimensions \cite{Celi2014,Zeng2015,Taddia2017}.
Furthermore, 1D superlattices have been realized with ultracold atoms \cite{Roati2008,Schreiber2015,Lohse2016,Nakajima2016}.
These 1D systems provide the main motivation for this research.

\subsection{Topological properties of the Harper-Hofstadter model}
\begin{figure}[tb] 
  \centering 
  \includegraphics[width=\columnwidth, clip]{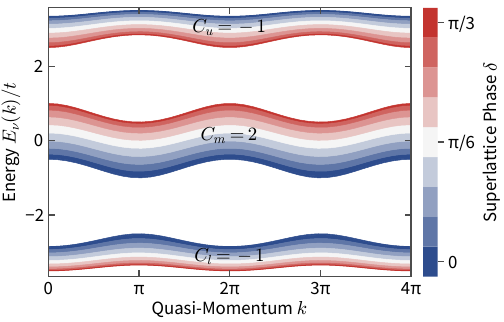} 
  \mycaption{(Color online) Topological bandstructure.}{
    Bandstructures for the noninteracting Fermi-Hubbard-Harper  model for $\alpha=1/3,\, V=3t$, as a function of $\delta$. 
    The bands correspond to the dispersion relation of the Harper-Hofstadter model, identifying $\delta$ with the transversal momentum $k_y$. 
Thus the Chern numbers describing the topology associated with a each set of bands share the topology of the bands of the 2D Harper-Hofstadter model, where
$C_{\{l,m,u\}} = \{-1,2,-1\}$ for lower, middle and upper band respectively.  
} 
\label{fig:bandstructure}
\end{figure}

The Harper-Hofstadter model hosts {topological} insulator phases, since its bands have nontrivial Chern numbers \cite{Thouless1982} (see \cref{fig:bandstructure}).
The Berry curvature is usually expressed in terms of antisymmetrized derivatives with respect to quasi-momentum $k_x,\,k_y$,
\begin{align}
    \label{eq:curvature_k}
    F(\vec k, \nu)\propto&\,\imag\left(\partial_{k_x}\bra{u_{\vec{k}, 
    \nu}}\right)\partial_{k_y}\ket{u_{\vec{k},\nu}} - 
    \left(k_x\leftrightarrow k_y\right)\,,\\
    C_\nu =&\,\frac{1}{2\pi}\int_{\rm BZ}\dd \vec k\, F(\vec k,\nu)\,,
\end{align} 
where $\ket{u_{\vec k,\nu}}$ are the eigenstates of the free 2D Hamiltonian for the $\nu$-th band.

This definition \cref{eq:curvature_k} of the Berry curvature $F$ can only be used in the noninteracting case, where quasi-momentum $\vec k$
is a good quantum number.
In our hybrid-space, one-dimensional model, we differentiate with respect to $k_x$ and $\delta$, using the 
replacement  $\partial_{k_y}\propto\partial_\delta$.

In the interacting case, the Berry curvature is generally defined on a family of Hamiltonians \cite{Avron1983,Simon1983}.
For a many-body system, we can introduce twisted boundary conditions, such that the Chern number is defined with respect to the twist angle $\theta$ \cite{Niu1985},
\begin{equation}
C\left(\{\ket{\psi}\}\right)=\frac{1}{\pi}\int_0^{2\pi}\dd \theta \int_0^{2\pi}\dd \delta\;
\Im\left[\left(\partial_\delta\bra{\psi}\right)\partial_\theta\ket{\psi}\right]\,,
\label{eq:berry_curvature}
\end{equation}
where $\ket\psi\equiv\ket{\psi(\delta,\theta)}$ is the {unique} many-body groundstate.

For $L\to \infty$, twisted boundaries do not affect bulk properties \cite{Niu1985, hastings2015, kudo2018}, and the family $\{\hat H(\delta,\theta)\}_{\mathbb{T}^2}$ 
 \cref{eq:fhh} can realize a topological charge pump: As $\delta$ is changed {adiabatically} to $\delta+2\pi$, a quantized amount of $C$ charges is transported through the system.

\subsection{Symmetries  \label{sec:symmetries}}

The Fermi-Hubbard-Harper model possesses several symmetries that can be exploited to  understand the groundstate physics and to improve the computational effort. 
These symmetries are used throughout the paper and we detail the relevant ones here. 

In the gauge chosen for \cref{eq:fhh}, it is obvious that $\hat H(\delta,\theta)$ is invariant under shifts by $2\pi$ of both twist angle $\theta$ and superlattice phase $\delta$.
However, for periodic boundary conditions, there is a higher symmetry related to the superlattice phase: 
shifting $\delta\to \delta + 2\pi /q$ with an additional gauge transformation merely corresponds to a {translation} $\hat T_{j'-j}$ of the
superlattice by $j'-j$ sites, defined through the modulo inverse,
\begin{equation}
    p(j'-j) = 1  \mod q\,,
\end{equation}
where $\alpha \equiv p/q$.
Thus, it suffices to compute groundstates for $\delta \in [0,2\pi/q)$.
Due to the gauge choice in \cref{eq:fhh}, i.e., complex hopping on one bond only, the Berry
curvature is not invariant under $\delta \to \delta+2\pi/q$. 
We need to construct states for $2\pi/q\leq \delta < 2\pi$ explicitly,
\begin{equation}
    \ket{\psi(\delta + 2\pi/q, \theta)}=\hat{T}_{j'-j}\eul^{-\imag \theta 
    \sum_{l=1}^{j-j'}\hat{n}_l}\ket{\psi(\delta,\theta)}\,.
    \label{eq:delta_translation}
\end{equation}
We choose $ j \leq j' < q$ 
and  $|\psi(\delta,\theta)\rangle$ is the groundstate for given values of $\delta$ and $\theta$.

The 2D Harper Hofstadter model is {particle-hole} symmetric around $E=0$. 
For the Fermi-Hubbard-Harper model of \cref{eq:fhh}, this symmetry is not present at any individual value of $\delta$. 
However, under a shift of $\delta \rightarrow \delta+\pi$, and $\theta \rightarrow -\theta$, the particle-hole symmetry is recovered. 

The interaction term in \cref{eq:fhh} also preserves the particle-hole symmetry: 
Exchanging $\Ch_{j,\sigma}\leftrightarrow \C_{j,\sigma}$ in \cref{eq:fhh}, we find a shifted superlattice $\delta \to \delta +\pi$ and reversed flux $\theta \to -\theta$.
Shifting $\theta$ does not change the Chern number as the curvature is integrated over the entire torus.
But changing the direction of the flux $\theta$ flips the sign of the Berry curvature in \cref{eq:berry_curvature} and thus changes the {many-body}
Chern number: $C(\rho, U)\to -C(2-\rho, U)$. 
Note that this implies $C(1, U) = -C(1, U) = 0$.
Because of the particle-hole symmetry,  it is sufficient to study phases and their topological properties for  $\rho \leq  1$. 

We further note that in the case of periodic boundary conditions, at values of $\delta = 2\pi n/3 $ and $\delta = 2\pi(1/6 + n/3) $ for $n \in \mathbb{Z}$, the system also possesses an inversion symmetry. 
The presence of an inversion symmetry allows for the existence of one-dimensional symmetry-protected topological states~\cite{Turner2011, Sirker2014}. 
These special $\delta$ points are important for the understanding of the possible topological properties of a family of Hamiltonians, $\{\hamil(\delta,\theta)\}_{\mathbb{T}^2}$. 
For $\delta =0, 2\pi/3$ and $4\pi/3$, there are no topological transitions at $\rho<1$, yet they occur at $\rho'=2 -\rho $ by particle-hole symmetry.

More specifically, a lattice-inversion symmetry constrains the many-body Zak phase which is defined as  
\begin{equation}
  \label{eq:zak-phase}
  \varphi = i\int_0^{2\pi}d\theta \braket{\psi(\delta, \theta)}{\partial_\theta \psi(\delta, \theta)}\,. 
\end{equation}
The Zak phase can have only two values $\varphi=0,\pi$ mod $2\pi$, differing by exactly $\pi$.
These two values of the Zak phase are topological invariants that cannot change under symmetry-preserving perturbations of the Hamiltonian without closing the many-body gap.

For open boundary conditions,
the choice of the unit cell can become important. 
This is typical of symmetry-protected topological states, where the choice of boundaries determines the presence or absence of gapless edge states \cite{asboth2016}. 
In our case, a choice of $\delta = \pi/3$ leads to an intra-cell site-centered symmetry, meaning that the lattice will retain its inversion symmetry. 
For $\delta = \pi $ and $\delta = 5\pi/3$ the lattice loses its inversion symmetry with open boundary conditions. 
We refer to these situations as symmetric and asymmetric lattice configurations, respectively.

\section{Methods \& observables}
\label{sec:methods}
\subsection{DMRG}
All numerical results presented in this paper were obtained using the DMRG algorithm \cite{White1992,Schollwoeck2011}.
We employ a single-site variant called DMRG3S \cite{Hubig2015}.
Particle-number conservation and $SU(2)$ symmetry in the spin sector with an associated quantum number $S$ of the model are 
exploited.
Thus, we fix the number of particles and the total spin for each computation.

In DMRG, the groundstate is represented as an open chain of rank-three tensors, a matrix-product-state (MPS) \cite{Schollwoeck2011}. 
Throughout this paper, we consider both periodic boundaries, required for the flux piercing, and open boundaries, which are numerically less challenging.

For periodic boundaries, we show data for up to $L=42$ sites, while calculations with open boundary conditions were performed up to $L =600$.
In both cases we used  $SU(2)$-reduced bond-dimensions up to $m=2000$ \cite{McCulloch2007}, which roughly corresponds to $m=5000$ when only using Abelian symmetries.

\subsection{Observables}
\label{sec:observables} 
As DMRG performs  a groundstate search, the energy of the state is
obtained in each step.
Furthermore, we estimate \cite{Hubig2018} the energy
variance $\mbox{var} (\hamil)= \langle \hat H^2 \rangle - \langle \hat H \rangle^2$ in order to quantify convergence. 
We can estimate the energy difference to the true groundstate $\delta_E:=E_{\rm DMRG}-E_0\approx\mathrm{var}(\hamil)/(E_1-E_0)$ using the energy difference of the two lowest states $E_{0}$ and $E_1$, assuming the DMRG wavefunction is a superposition of these two states.

This error estimate results from the following arguments. We assume that the DMRG wave function $| \psi\rangle_{\rm DMRG}$ is an 
overlap of the ground $|\psi_0\rangle$ and the first excited state $|\psi_1\rangle$  with $0<\alpha \lesssim 1$
\begin{equation}
| \psi\rangle_{\rm DMRG} = \alpha | \psi_0 \rangle + \sqrt{1-\alpha^2} | \psi_1 \rangle \,.
\end{equation}
Then:
\begin{equation}
\delta_E= \langle \psi_{\rm DMRG} | \hat H | \psi_{\rm DMRG} \rangle - E_0 = (1-\alpha^2)(E_1-E_0)
\end{equation}
The variance is:
\begin{equation}
  \begin{aligned}
\mbox{var}(\hat H) &=& \alpha^2 (1- \alpha^2) (E_1 -E_0)^2 \\
		   &\approx&  (1- \alpha^2) (E_1 -E_0)^2
\end{aligned}
\end{equation}
and therefore, $\delta_E \approx \mbox{var}(\hat H) / (E_1-E_0)$.
 
\subsubsection{Energy gaps \label{sec:observables-energies}}
Using DMRG, we compute groundstates in different particle number $N$ and SU(2) spin-symmetry
sectors to obtain different many-body gaps. 
Excitation gaps are crucial for obtaining quantum phase diagrams and for establishing  topological properties: The Chern number is defined only for a
groundstate manifold which is gapped everywhere.
{Topological} transitions, changing the Chern number, require degenerate
{points} on the groundstate manifold.
The  groundstate is always in the spin-singlet sector $S=0$, but we also
compute the lowest energy state in the spin-triplet sector $S=1$. Comparing these
energies, we can find different types of many-body gaps.

First, varying the particle number and keeping the total spin fixed gives the chemical
potential $\mu=\mu(N)$.
We search for incompressible states where $\partial \mu/\partial n \to \infty$, indicating 
insulating behavior. 

The charge gap is defined as 
\begin{eqnarray}
    \label{eq:charge_gap}
    \Delta_{\rm charge} (N) &=&\big[ E_0(N+2, S=0)  +E_0(N-2,S=0) \nonumber \\
	&&  -2E_0(N,S=0)\big]/2\,.
\end{eqnarray}  
Keeping the particle number constant, we define the {spin} gap between
singlet $S=0$ and triplet $S=1$ sector,
\begin{equation}
    \Delta_{\rm spin}(N) =  E_0(N, S=1)- E_0(N, S=0)\,.
     \label{eq:gapspin}
\end{equation}
We also compute the first excited state in the same symmetry sector, by orthogonalizing
both states during the DMRG run. Using $E_1$, we obtain the {internal} gap,
\begin{equation}
    \Delta_{\rm int}(N) = E_1(N, S=0)- E_0(N, S=0) \,.
    \label{eq:gap_int}
\end{equation}

We note that the internal gap need not be the smallest gap in the system, as states in other spin sectors may have lower energies. 

\subsubsection{Computing topological properties}
\label{sec:fukui_method} To compute the Berry curvature, we find groundstates for a
discretized grid on $\{\hamil(\delta,\theta)\}_{\mathbb{T}^2}$. 
As described in \cref{sec:symmetries}, with periodic boundary conditions, the system is symmetric under shifts in the superlattice phase: $\delta\rightarrow \delta + 2\pi n/3$.
Therefore, it is possible to relate the groundstate wavefunction found in the range $\delta\in[0,2\pi/q)$ to other states on the $\left\{\delta,\theta\right\}$-manifold via a translation and a gauge transformation. 

\begin{figure}[tb]
  \centering
  \includegraphics[width=\columnwidth, clip]{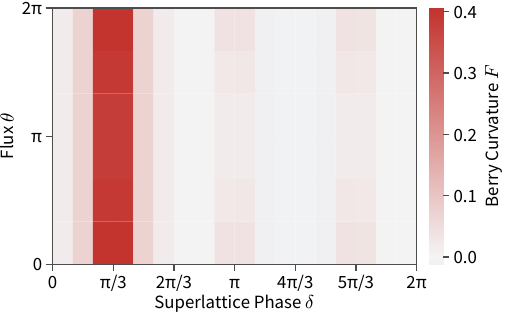}
  \mycaption{(Color online) Berry curvature in the strongly interacting regime.}{Computed from
  groundstates for $L=24$ and $V=3t$ at $U=12t$ and $\rho=2/3$. The integrated curvature thus yields a Chern number
$C(\rho,U)=1$. Only groundstates for $\delta \in [0, 2\pi/3)$ 
were calculated; all other overlaps were obtained exploiting the respective
symmetry, see \cref{sec:symmetries}. The curvature is nonzero close to
$\delta\in\{1,3,5\}\pi/3$. Note that the curvature is not invariant under shifts 
$\delta \to \delta + 2\pi/3$ due to the inhomogeneous gauge choice, see
\cref{sec:symmetries}.
}
\label{fig:berry_curvature}
\end{figure}

We compute overlaps of all groundstates adjacent on the $\left\{\delta,\theta\right\}$ discretized torus and evaluate the Berry curvature using the
 method by \citet{Fukui2005}:
\begin{eqnarray}
    \label{eq:fukui_curvature}
    F(\delta,\theta)&=&\,\Im\ln\frac{
      \braket{\psi(\delta, \theta)}{\psi(\delta', \theta)}\braket{\psi(\delta', \theta)}{\psi(\delta', \theta')}}
      {\braket{\psi(\delta, \theta')}{\psi(\delta', \theta')}\braket{\psi(\delta, \theta)}{\psi(\delta, \theta')}}\, ,\\
    && \delta'=\,\delta+\Delta_{\delta}\,,\quad \theta'=\theta+\Delta_\theta\,,
\nonumber 
\end{eqnarray}
where $\Delta_{\delta,\theta}$ is the grid spacing in parameter space.
This expression can be understood as the Berry phase gained when moving along a closed loop in the $\{\delta,\,\theta\}$-parameter space. 
Overlaps can easily be computed from MPS representations \cite{Schollwoeck2005}. 
The Berry curvature for one configuration in the strongly correlated insulator is plotted in \cref{fig:berry_curvature}.

By construction, the numerically computed Chern number,
\begin{equation}
    C(\rho ,U) = \frac{1}{2\pi}\sum_{\delta,\theta\in[0,2\pi)}F(\delta,\theta,\rho ,U)\,,
\end{equation}
is always an integer. 
This remains true even if the grid spacing is too coarse or the groundstate manifold is not gapped everywhere. 
However, in these cases the computed Chern number will not necessarily be stable under small changes. 

We verify the degree of convergence by using the real part of the logarithm in \cref{eq:fukui_curvature}. 
This measure is small when the overlap between wavefunctions at adjacent points on the grid  is close to unity.
We choose a  $6\times 6$ grid for $\delta\in[0,2\pi/q),\,\theta \in[0,2\pi)$ such that the absolute values of all overlaps are large, except when the internal gap is zero.

Furthermore, when studying open boundary conditions, we compute the center-of-mass coordinate $\overline X(\delta)$, which can easily 
be computed using DMRG and can be measured in experiments \cite{Lohse2016,Nakajima2016}.
The definition of $\overline X(\delta)$ is:
\begin{equation}
  \overline X(\delta) =\frac{1}{L}\sum_j j \bra{\psi(\delta)}\N_j\ket{\psi(\delta)}\,.
    \label{eq:cm_motion}
\end{equation}
Here, $|\psi(\delta) \rangle$ is the groundstate at a given $\delta$.
Note that a flux $\theta$ is merely a static gauge transformation for open boundaries, and thus does not matter here.
As we compute groundstates for each value of $\delta$ independently, 
there is no accumulation of charge at either end during a pump cycle: the center-of-mass coordinate returns to its initial value as $\delta\to\delta+2\pi$. 
Instead, quantized charge
transport can be observed as {discontinuities} of the change of the center-of-mass coordinate. 
These discontinuities correspond to the shift of an occupied edge mode from one side of the system to the other \cite{Hatsugai2016}.

\subsubsection{Additional observables}
For each groundstate, we compute the  one-particle (reduced) density-matrix (OPDM), containing
all (normal) single-particle observables.  Since we enforce $SU(2)$ spin-symmetry, we can only
compute the spin-independent OPDM $\rho^{(1)}$,
\begin{equation}
    \rho^{(1)}_{i,j} = \sum_{\sigma} \left\langle \Ch_{i,\sigma}\C_{j,\sigma} \right\rangle\,.
\end{equation}
We exploit translational symmetry to restrict one index to $0\leq i<q$ for nondegenerate states.
From the OPDM, we can extract occupations of eigenstates of the free Hamiltonian, $U=0$. 
The noninteracting eigenbasis is obtained from a  linear transformation $\C_{\nu,\tilde
k,\sigma}=\sum_j \big(\vec{a}_{\nu, \tilde k}\big)_j \C_{j,\sigma}$. 
For nondegenerate states, we can thus compute total
occupation $n_\nu$ of the $\nu$-th bands
\begin{equation}
    n_\nu \equiv \sum_{\tilde k}\langle\N_{\nu,\tilde k}\rangle =
    \sum_{i,j} \big(\vec a_{\nu,\tilde k}\big)_i^\dagger \rho^{(1)}_{i,j}\big(\vec a_{\nu,\tilde k}\big)_j \,,
   \label{eq:momentum_dist}
\end{equation}
where $ \langle\N_{\nu,\tilde k}\rangle = \sum_\sigma \langle\N_{\nu,\tilde k, \sigma}\rangle$ and $ \langle\N_{\nu,\tilde k, \sigma}\rangle$ is the quasi-momentum distribution function of fermions with spin $\sigma$ in the $\nu$-th band, with $\nu \in \{l, m, u\}$ corresponding to the lower, middle and upper band, respectively.

In addition to real-space occupation numbers $\langle\N_j\rangle=\rho^{(1)}_{j,j}$, we compute
expectation values for local single $\hat P^{(1)}_{j}$ and double occupation and $\hat P^{(2)}_{j}$, respectively, related via $\hat n_j=\hat P^{(1)}_{j}+2\hat P^{(2)}_{j}$. 
Here, $\hat P^{(i)}_{j}$ is the projector onto the manifold with $i$ particles at the  $j$-th site. 

To identify the spontaneously dimerized phase, we compute the bond-order parameter:
\begin{equation}
  \langle \hat B\rangle = \frac{1}{L/3}\sum_{j=0, \sigma}^{L/3-1}\langle \Ch_{3j,\sigma} \C_{3j+1,\sigma} - \Ch_{3j+1,\sigma}\C_{3j+2,\sigma} + \hc \rangle\,.
\end{equation}
For our case of a three-site superlattice ($q=3$) with phase $\delta=\pi/3$ according to \cref{fig:realspace_model}, site $0$ and $2$ are energetically degenerate while site $1$ is lower in energy. 

\section{Quantum Phases of the 1D Fermi-Hubbard-Harper model}
\label{sec:groundstates} 
In this section, we discuss quantum phases of groundstates for individual
$\hamil(\delta,\theta)$ from the family of Hamiltonians $\{\hamil(\delta,\theta)\}_{\mathbb{T}^2}$.
As the twist angle $\theta$ is a boundary effect, it does not affect bulk physics in the thermodynamic limit \cite{hastings2015, kudo2018}.
However, the superlattice phase $\delta$ can affect quantum phases. 
For example, the SDI phase (to be discussed below) appears only for certain values of $\delta$, related to the lattice-inversion symmetry discussed in \cref{sec:symmetries}. 

Quantized charge transport in the family of Hamiltonians can only occur if the physical
state is insulating for the entire pump cycle \cite{Thouless1983, Niu1985}. 
Conversely, the Chern number of the manifold of groundstates can only change when the many-body gap closes.
We thus start by studying insulating phases.

In the following sections, we restrict ourselves to a three-site superlattice,
$\alpha\equiv p/q=1/3$. For this configuration, there are three separated energy bands,
which are all topologically nontrivial, see \cref{fig:bandstructure}. Furthermore, we
will choose  $V=3t$ as the strength of the potential unless stated otherwise. For this value of $V/t$,
the band gaps are comparable to the hopping matrix elements $t$. 
We find that significantly
stronger superlattice potentials do not change the physical behavior qualitatively.

\begin{figure}[tb]
    \centering
    \includegraphics[width=\columnwidth, clip]{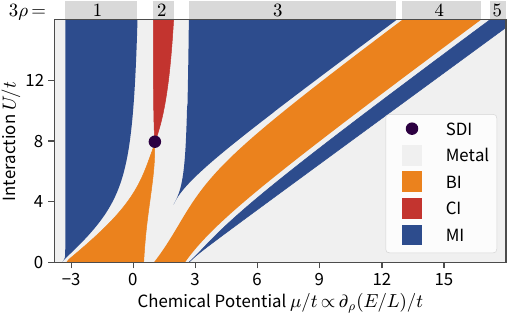}
    \mycaption{(Color online) Grandcanonical phase diagram.}{Groundstate phases of the Fermi-Hubbard-Harper model
    with $\alpha=1/3$, computed for $\delta=\pi/3$ and $V=3t$ with open
    boundaries.
    For $\rho\in\{2,4\}/3$ there are band insulators (BI) for $U=0$.
    At $U\approx8t,\,\rho=2/3$ there is a transition to a spontaneously dimerized insulator (SDI) and then to a strongly interacting correlated insulator (CI).
    Mott insulators (MI) appear for half-filled bands, $\rho=1/3, 1$, and $5/3$.
    The phase boundaries $\mu(U, \rho\pm 1/L)$ are extrapolated in $L\to \infty$.
    The density $\rho$ in each of the incompressible phases is indicated on the top of the figure.}
\label{fig:phase_diag}
\end{figure}

\subsection{Grandcanonical phase diagram}

In order to obtain  the phase diagram for the Fermi-Hubbard-Harper model we compute groundstates for
various particle numbers and interaction strengths $U$.  
As described in \cref{sec:observables-energies}, we can infer the chemical potential $\mu$ and the $\{\mu,U\}$-phase
diagram.

A phase diagram obtained from DMRG data for  open boundaries
is shown in \cref{fig:phase_diag} for
$\delta=\pi/3$.  The analysis of the  charge gap [see Eq.~\eqref{eq:charge_gap}]
suggests the existence of  insulating phases for five different fillings.  At $\rho=2/3$ and $4/3$, there are band
insulators, already present without interactions at  $U=0$.  Furthermore, for half-filled
bands, $\rho=1/3,1$ and $5/3$, Mott-insulating phases emerge for $U>0$.

An interesting sequence of phases exists at filling $\rho=2/3$: upon increasing $U/J$, the BI ultimately (via two
transitions) turns into a correlated insulator at $U\gg J$. We use the term CI to distinguish this large $U/J$ phase from 
MIs since at filling  $\rho=2/3$, the bands are either empty or filled. The term CI is also used in parts of the  literature
in the same context \cite{Manmana2004}. 
We  find evidence (see Sec.~\ref{sec:SDI}) that the intermediate phase is  a bond-ordered spontaneously dimerized insulating (SDI) phase separating the BI and CI phases at density $\rho=2/3$, 
indicated in \cref{fig:phase_diag}.

\subsection{Mott insulator at density \texorpdfstring{$\rho=1/3$}{rho=1/3}}
\label{sec:mi_1_3} For particle density $\rho=1/3$ and without interactions $U$, the lowest band
is half-filled, and we are in a metallic phase for all $\delta, \,\theta$. As we saw in
\cref{fig:phase_diag}, a charge gap opens for weak interactions and the phase appears to be 
a Mott insulator for all $U>0$.

While the charge gap [see Eq.~\eqref{eq:charge_gap}] is comparable to the size of the gaps for the
band-insulating phases in \cref{fig:phase_diag}, there can be gapless spin excitations for
the infinite system in this Mott insulator (see the discussion in \cref{sec:scl}).

\begin{figure}[tb]
    \centering
    \includegraphics[width=\columnwidth, clip]{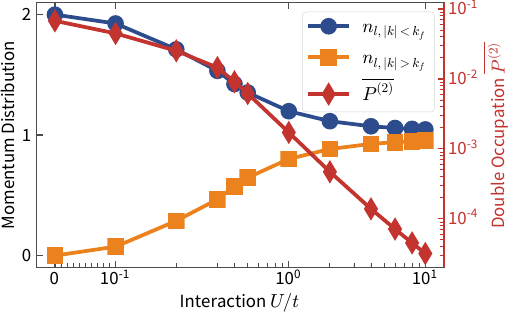}
    \mycaption{(Color online) Partially integrated momentum-distribution function and real-space double occupation
    for $\rho=1/3$.}{
    We show data for superlattice amplitude $V/t=3$ and phase $\delta=\pi/3$.
    $\overline{P^{(2)}}$  is the double occupation averaged over one superlattice cell. 
    Increasing the strength $U$ of the interaction suppresses double occupation for the MI phase.
    In momentum space, all particles remain in the lowest band,
    but, as $U$ increases, states below and above the Fermi wavenumber $k_F$  become evenly
    occupied. The lines are guides to the eye.
}
\label{fig:mott_occ}
\end{figure}

Increasing onsite repulsion $U/t$ obviously suppresses the double occupation $\langle\hat P^{(2)}_{j}\rangle$ on all lattice sites $j$. 
In \cref{fig:mott_occ}, we illustrate that real-space double occupation decreases with $U/t$.
Moreover, also  in momentum space, occupation numbers change as a function of 
increasing interaction strength.
As shown in the same figure, for $U=0$, the lower half of the lowest band is fully occupied by each spin species.
When interactions increase (in the range considered in Fig.~\ref{fig:mott_occ}, the   particles mostly remain in the lowest band of the noninteracting model
, but we approach a constant momentum-distribution function $\expect{\N_{\nu=l,k}}=1$ for the entire lowest band.  
Only considering single-particle observables $\Ch_{i,\sigma}\C_{j,\sigma}$, such as particle number $\hat{P}^{(1)}_{j}$ or the momentum-distribution function $\expect{\N_{\nu,k}}$, the system behaves much like free,
spinless fermions at the same particle density: 
For a single species of fermions, double occupation is prohibited by Pauli's exclusion principle
and, without interactions, the lowest band would be singly filled at density $\rho=1/3$.

\subsubsection{Strong-coupling limit}
\label{sec:scl}

In order to understand the phases present in our model, we  use
Schrieffer-Wolff (SW) perturbation theory to simplify the problem in certain
limits. In particular, SW theory allows us to understand the effective
spin-sector behavior typical of Mott insulators when there is a significant
charge gap.

For the single-band one-dimensional Hubbard model at half-filling \mbox{($\rho=1$)}, any nonzero
Hubbard interaction induces a charge gap \cite{Essler2005}. In the limit of $t \ll U$,  the Hubbard
interaction projects out doubly-occupied sites, as these sites have energy $U$. The groundstate therefore lives in the manifold of singly-occupied sites.

One can then use Schrieffer-Wolff perturbation theory to derive an effective Hamiltonian which describes the low-energy physics in this manifold of states \cite{Essler2005}:
\begin{equation}
  \label{eq:XXZ}
  \hamil^{\rm eff}_S = J\sum_{i} \hat{\mathbf{S}}_i \cdot \hat{\mathbf{S}}_{i+1} + \mathcal{O}\left(t^2/U^2\right)\, . 
\end{equation}
Here, $\hat{\mathbf{S}}_i$ labels the spin-$1/2$ degree of freedom on site $i$ and $J = t^2/U$ is the induced spin-spin interaction. 
This effective model is the well-known isotropic Heisenberg chain, which has gapless spin
excitations \cite{schollwoeck2008quantum}. 
Importantly, as the groundstate manifold and the original Hamiltonian have a global $SU(2)$ symmetry, the effective Hamiltonian will also contain only $SU(2)$ invariant terms. 

In the present case of a  model with a lower degree of translational symmetry and away from half filling, the effective model is more complicated. 
Following~\cite{Torio2006}, we can write the effective strong-coupling model as follows: 
\begin{eqnarray}
  \hamil = & - t \sum_{i, \sigma} \hat P \left( \Ch_{i + 1, \sigma} \C_{i, \sigma} + \mathrm {H.c.} \right) \hat P + \sum_{i} \Delta_{i} \N_{i} \nonumber \\
  & + \sum_{i, \delta = \pm 1} t_{i} ^ {\rm ch} \hat P \left[ \Ch_{i + \delta, \sigma} \C_{i - \delta, \sigma} \left( 2 \hat{\mathbf {S}}_{i} \cdot \hat{\mathbf {S}}_{i - \delta} - \frac {1} {2} \N_{i} \right) \right] \hat P\nonumber \\ 
  & + \sum_{i} J_{i} \left( \hat{\mathbf {S}}_{i} \cdot \hat{\mathbf {S}}_{i + 1} - \frac {1} {4} \N_{i} \N_{i + 1} \right)   \label{eq:effective-tj}\,. 
\end{eqnarray}
Here, $\hat P = \prod _ { i } \left( 1 - \N_{ i, \uparrow } \N_{ i, \downarrow } \right)$ projects out all doubly-occupies sites, $\hat{\mathbf{S}}_i = \sum_{\alpha,\beta}\Ch_{i,\alpha}\mathbf{\sigma}_{\alpha \beta}\C_{i,\beta}$ are the spin operators, and the derived coupling strengths are:
\begin{equation}
\begin{aligned}
  J _ { i } & = \frac { 4 t ^ { 2 } U } { U ^ { 2 } - \Delta_{i,i+1} ^ { 2 } }\,, \\ 
  t _ { i } ^ {\rm ch} & =\frac{1}{2}\left(
    \frac { t ^ { 2 } } { U + \Delta_{i, i+1} } +
    \frac { t ^ { 2 } } { U - \Delta_{i, i-1} }
  \right)\,,
\end{aligned}
\end{equation}
where $\Delta_{i,j} = V\cos(2\pi j/3 + \delta) - V\cos(2\pi i/3 + \delta)$ is the potential difference between sites $i$ and $j$. 
Terms of higher order in $t/U$ have been omitted. This model describes a generalized $t-J$ model, which reduces to the homogeneous case when $V=0$.

The strong-coupling limit ($U \gg \Delta,\, t$) can be studied by solving first for the distribution of the charge degrees of freedom, and then treating the terms proportional to $J_i$ and $t_i^{\rm ch}$ perturbatively. 
This charge distribution can determined by finding the groundstate of a system of noninteracting spinless fermions $c^\dagger_{i}$ on the lattice in question. 
The effective spin Hamiltonian is then obtained  by projecting the Hamiltonian in \cref{eq:effective-tj} onto the charge distribution. 

For $\rho=1/3$, the charge distribution is the same as that of a system with a filled lowest band of the noninteracting spinless model. 
One  then recovers the effective model: 
\begin{equation}
  \hamil^{\mathrm{eff}} = \frac{1}{2}\sum_{i,j} J_{i,j}^{ \mathrm { \mathrm{eff} } } \left( \mathbf {\hat{S}}_{i}\cdot\mathbf{\hat{S}}_{j} - \frac{1}{4} \right),
\end{equation}
where
\begin{equation}
  \begin{aligned}
    J _{i,i+1} ^ { \mathrm { eff } } \approx 
    & \frac{1}{N} \sum_{i=0}^{N-1} \bigg[ 
    J_i \left\langle \hat{n}_i \hat{n}_{i+1} \right\rangle^\prime 
  + 2t_i^{\rm ch} \left\langle \hat{c}_{i-1}^\dagger \hat{c}_{i+1} \hat{n}_{i} \right\rangle^\prime  \\
& + 2t_{i+1}^{\rm ch} \left\langle \hat{c}_{i+2}^\dagger \hat{c}_{i} \hat{n}_{i+1} \right\rangle^\prime  \bigg] .\end{aligned}
\end{equation}
The expectation values $\langle \cdot \rangle'$ are taken with respect to the spinless-fermion background \cite{Torio2006}.
This model is a ``squeezed" Heisenberg chain, where the empty sites have been eliminated, and 
the $\hat{\mathbf{S}}_i$ refer to the spins attached to the $i$th fermion, which will be centered at the $i$th unit cell on average. This chain  therefore has length $N = \rho L$.
 
The spin chain inherits symmetries from the underlying lattice and the charge distribution. 
For $\rho=1/3$, this implies that the $J^{\mathrm{eff}}_i$ are homogeneous, and we recover the standard Heisenberg model, with one spin per unit cell. 
The effective spin model therefore predicts that the system has gapless spin excitations in the strong coupling limit, which is consistent with our numerical data. Note that this result is independent of $\delta$ and $\theta$. 

\subsection{Insulators at density \texorpdfstring{$\rho=2/3$}{rho=2/3}}

\begin{figure}[tb]
    \centering
  \includegraphics[width=\columnwidth, clip]{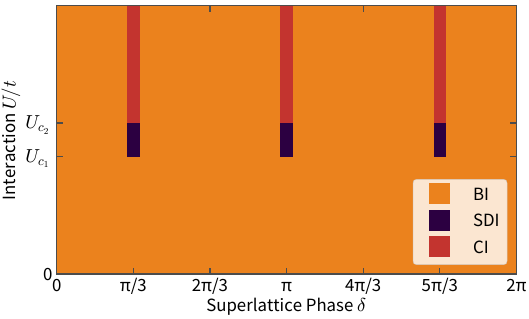}
  \mycaption{(Color online) Phase diagram for the $\rho=2/3$ insulator.}{The system, throughout the $\left\{\delta,U\right\}$-plane, is adiabatically connected to a band insulator at $U=0$. 
    However, along three lines at $\delta = \pi/3$, $\pi$, and $5\pi/3$, the lattice has a symmetry in the superlattice potential which leads to the phase structure outlined in \cref{fig:transition-sketch}.
  \label{fig:rho23-phase-diagram}}
\end{figure}
The system at density $\rho=2/3$ has a complicated phase diagram with several transitions.
The basic structure of the phase diagram is illustrated in \cref{fig:rho23-phase-diagram}.
In this section, we discuss the sequence of phases. 
To give a brief overview, at $U=0$ the system is a band insulator with a filled lower band. 
Apart from the symmetric lines along $\delta = \pi/3,\pi$ and $5\pi/3$, the BI survives at all $U/t$ and becomes strongly correlated as $U/t$ increases. 

\begin{figure}[tb]
    \centering
    \includegraphics[width=\columnwidth, clip]{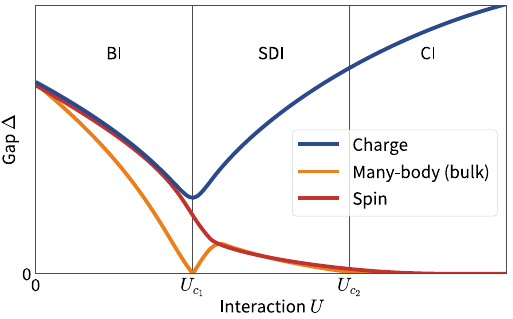}
    \mycaption{(Color online) Sketch of the phase diagram for  $\delta=\pi/3$ at $\rho=2/3$.}
    { Along this line, there are two  critical values of $U$, $U_{c_{1}}$ and $U_{c_{2}}$. 
$U_{c_{1}}$ marks an Ising-like transition between a band insulator (BI) and a spontaneously dimerized insulator (SDI) \cite{Lou2003,Manmana2004,Torio2006,Tincani2009}. 
The second critical value $U_{c_{2}}$ is the point of  a BKT-like transition between the SDI and a correlated insulator (CI). 
We also show the behavior of various energy gaps in these different phases for various types of excitations. 
The band insulator is completely gapped throughout. 
 The global many-body gap is identical to the  internal gap in the BI and closes at $U_{c_1}$ when the groundstate becomes doubly degenerate in the SDI, while the charge and spin excitations remain finite. 
The transition between the SDI and CI at $U_{c_2}$ occurs when the spin gap closes.
\label{fig:transition-sketch}}
\end{figure}

Along the symmetric lines, there are two phase transitions, which are sketched in \cref{fig:transition-sketch}. First, at $U_{c_1}$ the 
system undergoes  an Ising-like transition from the BI to a doubly degenerate SDI \cite{Torio2001,Lou2003, Manmana2004, Torio2006, Tincani2009}.
At a larger interaction strength $U_{c_2}$, there is a second transition of the Berezinskii-Kosterlitz-Thouless (BKT) type 
to a CI with gapless spin excitations \cite{Takada2001,Torio2001,Lou2003, Aligia2004, Manmana2004, Torio2006, Tincani2009}. 

\subsubsection{Band insulator}
At fermion density $\rho=2/3$ there is a charge gap at
$U=0$ (see Fig.~\ref{fig:phase_diag}), corresponding to the band gap of the Hofstadter model, see \cref{fig:bandstructure}.
We find that even strong interactions preserve the properties of the band-insulating phase for most parameters $\delta$ of the family \cref{eq:fhh}. 
 This band insulator is adiabatically connected to all points in the $\lbrace U,\delta \rbrace$-parameter space, except for the lines with the SDI and the gapless CI phases, as sketched in \cref{fig:rho23-phase-diagram}. 
While there is no phase transition throughout this region (except at $\delta=\pi/3,\pi, 5\pi/3$), there is a smooth change to a gapped strongly correlated state as $U/t$ increases.

\subsubsection{Spontaneously dimerized insulating phase}
\label{sec:SDI}

At $\delta = 2\pi/6 + 2\pi n/3$, between the CI phase and the BI phase, there is an intermediate bond-ordered phase, typical of ionic Hubbard models \cite{egami1993,fabrizio1999,Torio2001,kampf2003,Lou2003, batista2004, Manmana2004, otsuka2005, lombardo2006, kancharla2006, paris2007, byczuk2009, Tincani2009, garg2014, Loida2017}. 
This phase has been studied by a mapping to an exactly solvable  SU(3) antiferromagnetic Heisenberg chain \cite{batista2004} where the state was found to have both spin and charge dimerization. 
The dimerization spontaneously breaks the lattice-inversion symmetry which occurs at these values of $\delta$.

\begin{figure}[tb]
  \centering
  \includegraphics[width=\columnwidth, clip]{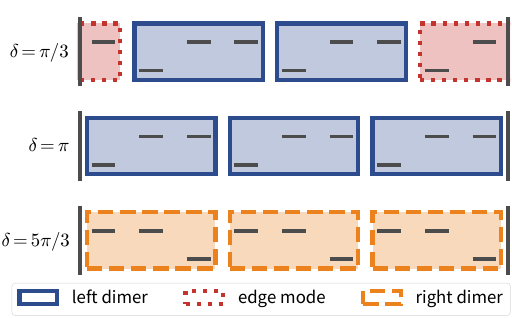}
  \mycaption{(Color online) Lattice configurations for open boundary conditions.}{Different choices of $\delta$ change the boundary conditions. 
    These boundary conditions determine edge-state properties of the groundstate of the SDI phase: The two asymmetric lattice configurations, $\delta = \pi$ and $\delta = 5\pi/3$, support left and right dimerized groundstates, respectively. 
    However, for $\delta= \pi/3$, the symmetric lattice configurations support neither groundstate, leading to the existence of gapless edge modes.} 
  \label{fig:edge-states}
\end{figure}

The situation is illustrated in \cref{fig:edge-states}. 
The choice of boundary conditions is particularly relevant for the SDI phase, where the symmetric configuration at $\delta = \pi/3$ splits a dimer, leading directly to the existence of gapless edge modes, and the asymmetric configurations each support one of the two SDI groundstates. 

A numerical study of the two-site ionic Hubbard model \cite{Manmana2004} suggests that bulk many-body gap and spin gap close at different interaction strengths, indicating the two-step sequence of phase transitions, while the charge gap does not close at any point. 
The case of a three site unit-cell, relevant to the present case of $\alpha\equiv p/q=1/3$ was studied in both \cite{Torio2006} and \cite{murcia2016}, where the same situation was found.

\begin{figure}[t]
    \centering
    \includegraphics[width=\columnwidth, clip]{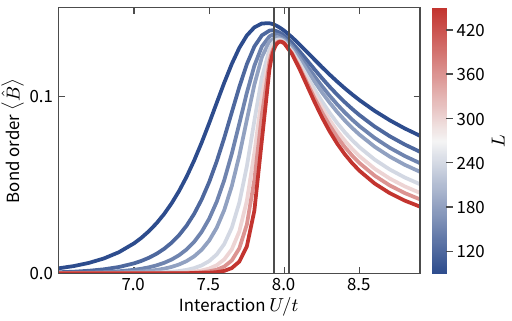}
    \mycaption{Bond-order parameter across the BI-SDI-CI transition.}
    {At $\delta=\pi/3$, the system  is as depicted in \cref{fig:realspace_model}, i.e., the potential energies on sites 1 and 3 are degenerate, and site 2 is lower in energy.
    Data were computed for open boundaries {without} inversion symmetry for $L\in\{90,120,150,180,240,300,360,450\}$. The vertical lines indicate the two phase transitions, as determined from the gap data shown in \cref{fig:gaps_scaled}.}
    \label{fig:bond_order}
\end{figure}

For a system with periodic boundary conditions, the BI and CI phases preserve the lattice-inversion symmetry, implying that the bond-order parameter vanishes, i.e., $\expect{\hat{B}}=0$. 
However, in the SDI, the lattice symmetry is spontaneously broken, leading to a doubly degenerate groundstate, and a finite value for $\expect{\hat{B}}$.
In \cref{fig:bond_order}, we show the bond-order parameter $\expect{\hat{B}}$ as a function of $U$ for open boundary conditions. 
The finite length of the system leads to a nonzero $\expect{\hat{B}}$ in the BI and CI phases, but $\expect{\hat{B}}$  disappears in the thermodynamic limit \cite{Manmana2004, Torio2006}. 
However, the appearance of the SDI phase is consistent with our data for large, but finite, system sizes. 
The precise effects of the open boundary conditions and the relationship to lattice symmetries is subtle and is discussed in more detail in \cref{sec:edge-symm-appendix}.

\begin{figure}[tb]
    \centering
    \includegraphics[width=\columnwidth, clip]{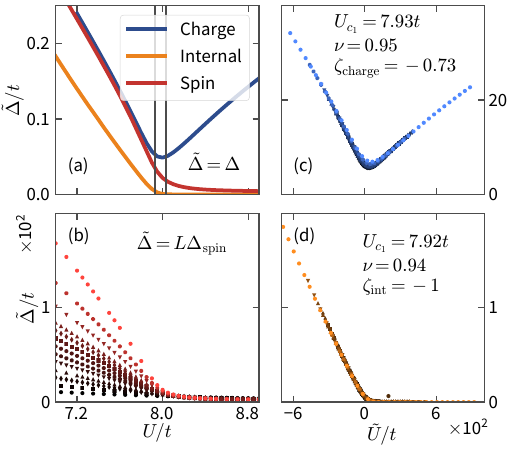}
    \mycaption{Gaps at the phase transitions for $\delta=\pi/3$.}{All data were computed with open boundaries. 
        (a) Finite-size gaps $\Delta$ versus interaction strength $U$ for $L=600$. 
        (b) Spin gap, divided by chain length $L$ such that the data should collapse above the BKT transition.
        (c) \& (d) Rescaled data for (c) charge and (d) internal gap computed for $L=30$ (dark) to $L=600$ (bright) ($L\in\{30, 60, 90, 120, 150, 180, 210, 240, 270, 300, 360, 450, 600\}$).
         In  (c) and (d),   we present a scaling collapse using: 
        $\tilde U=L^{1/\nu}(U-U_c)$, $\tilde{\Delta} = L^{-\zeta/\nu}\Delta$ with different parameters $\nu,\zeta,U_c$, estimated from the finite-size data.
}
    \label{fig:gaps_scaled}
\end{figure}

We show finite-size data in \cref{fig:gaps_scaled}(a) for all three gaps defined in Eqs.~\eqref{eq:charge_gap}, \eqref{eq:gapspin}, and \eqref{eq:gap_int}.
Data were obtained with open boundary conditions and $L= 600$ for superlattice phase $\delta=\pi/3$.
The charge gap exhibits a minimum at $U/t \approx 8$ while spin and internal gap decrease  monotonously with $U$.
This behavior is suggestive of a vanishing of the spin gap at large $U/t$ and a zero of the internal gap
at a lower critical value of $U$, which we will further substantiate below. 
The fact that the internal gap becomes very small  for $U/t\gtrsim 8$ is due to degenerate edge modes (see \cref{sec:edge-symm-appendix}).

Quantitatively, we determine exponents and critical values of the interaction strength for the first transition from a scaling collapse of the charge and internal gap in Figs.~\ref{fig:gaps_scaled}(c) and (d):
we assume these scaling relations:
\begin{equation}
\tilde{U}_r = L^{1/\nu} (U- U_{r})\,, \quad  \tilde{\Delta}_r = L^{-\zeta_r/\nu}\Delta_r\,.
\label{eq:scaling}
\end{equation} 
Here, $r\in\{{\rm int, charge}\}$ labels the gaps and the critical $U$, with $U_{\rm int} = U_{\rm charge} =  U_{c_1}$.
We find general agreement between the  data for the excitation gaps shown in \cref{fig:gaps_scaled} and the data for $\expect{\hat{B}}=0$ shown  in \cref{fig:bond_order} regarding
the extent of the SDI phase.
The scaling collapse of both gaps leads to  the same value for critical interaction strength $U_{c_1}$ related to the first transition and the exponent $\nu\approx 1$ matches an Ising transition (see, e.g., \cite{Manmana2004,Tincani2009}).

The spin gap should scale according  to the BKT universality class at the transition from SDI to CI and therefore,  
$L\Delta_{\rm spin}$ is expected to become independent of $L$ at the phase transition.
The data shown in \cref{fig:gaps_scaled}(b) is consistent with a BKT transition at some  $U_{c_{2}}/t \gtrsim 8$.
We estimate the critical interaction strength $U_{c_{2}}$ using a (conventional, non-BKT) scaling collapse shown in \cref{fig:spin_collapse}
and obtain $U_{c_{2}}/t \approx 8.03$.

\subsubsection{Correlated insulator}
\label{sec:MI_2_3}
Only for particular values of the superlattice phase $\delta\in\{1,3,5\}\pi/3$, we observe transitions to correlated insulating phases.
For these values of $\delta$, the model corresponds to the AB$_2$ ionic Hubbard chain \cite{murcia2016}. 
This lattice configuration is sketched in \cref{fig:realspace_model}: 
Two lattice sites are energetically degenerate, while the third site is lower in energy. 
In an `atomic picture' (i.e., $t\to 0$), for density $\rho=2/3$, we would expect different states for small and large interaction strength $U$ compared to the superlattice potential $V$: 
If $U$ is weak, there are two particles localized in the site of the lowest
energy. 
Strong repulsion $U\gg V$ prohibits double occupation, and therefore, there is only one particle in the potential minimum, while the other particle is delocalized over the remaining sites. 

Assuming the atomic limit $t\ll V$, we can relate real-space and band occupations.
When we choose a homogeneous gauge, quasi-momentum $k$ is a conserved quantity for the noninteracting Hamiltonian \cref{eq:fhh}. 
Thus, we can express it in momentum space, using a vector of $q=3$ creation operators $\VC_{k, \sigma}$ for each spin $\sigma$ and momentum $k$,
\begin{align}
    \hamil=&\sum_{k, \sigma} \VCh_{k, \sigma} h_k \VC_{k, \sigma}\,,\\
    h_{k}=&\left(\begin{matrix} V \cos(\delta) & -t& -t\eul^{\imag (k-3\theta/L)}\\
        -t & V\cos\left(\frac{2\pi}{3}+\delta\right) & -t \\ 
        -t\eul^{-\imag (k-3\theta /L)} & -t & V \cos\left(\frac{4\pi}{3} +\delta\right)
    \end{matrix}\right)\,. \nonumber
\end{align}

This $q\times q$ matrix becomes diagonal for strong potentials $t/V\to 0$. 
In this limit, the states of each band are supported on only one lattice site in each superlattice cell.
Therefore, we should expect that, given a strong potential $V$, the interaction does not only
suppress double occupation in real space, but also in momentum space.

\begin{figure}[tb]
    \centering
    \includegraphics[width=\columnwidth, clip]{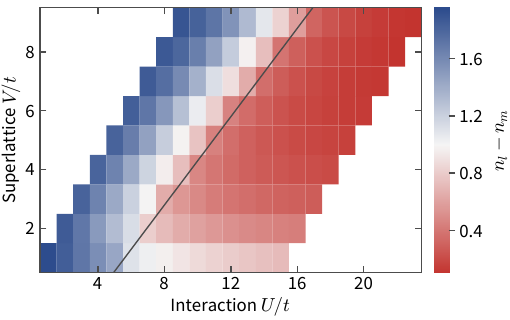}
    \mycaption{(Color online) Band occupation for $\rho=2/3$.}{We show the band insulator--to--correlated-insulator transition with the lowest band filled, $\rho=2/3$, for $\delta=\pi/3,\,\theta=0$. 
      For a given superlattice potential strength $V$, particles are transferred to the middle band as the interaction strength increases. 
      The occupation of the upper band remains small ($n_{u} \lesssim 0.1$) everywhere. 
      The gray solid line indicates the topological phase transition where the Chern number changes, as in \cref{fig:top_transition}. 
    Calculations were performed for $L=18$.}
    \label{fig:uv_band_occ}
\end{figure}

We show the density difference for the band occupation of the lower two bands for different $U$ and $V$ in \cref{fig:uv_band_occ}. 
For a sufficiently large potential strength $V/t\gtrsim 3$ 
our argument seems to hold and double occupation of bands is suppressed monotonically by increasing  $U/V$. 
In the large $U$ limit, we find that the two lowest bands are occupied evenly. 
This corresponds to the charge density of a spinless-fermion model with the same density.
In the atomic limit, where $U,V \gg t$, we can estimate the location of the crossover to occur at $V= 3U/2$, where the double occupancy becomes energetically unfavorable for increasing $U$.
 
\subsubsection{Strong-coupling limit}
\label{sec:strong_coupling2_3}
We can again study the spin sector in the strong-coupling limit by performing Schrieffer-Wolff perturbation theory, as in \cref{sec:scl}. 
For $\rho=2/3$, we recover an effective spin chain of length $ L_{\rm eff}=L\rho =  N$.  
This also implies that the effective spin model has a unit cell of two spins, which we label $A$ and $B$.
The spin physics is then governed by the effective Hamiltonian:
\begin{equation}
  \label{eq:XXZ-dimer}
  \hamil^{\rm eff}_S = \sum_i \left\lbrack J \hat{\mathbf{S}}_{i,A} \cdot \hat{\mathbf{S}}_{i,B}  
  + J' \hat{\mathbf{S}}_{i,B} \cdot \hat{\mathbf{S}}_{i+1,A} \right \rbrack\,.
\end{equation}
Here, the $\hat{\mathbf{S}}_{i,A(B)}$ labels the $A$($B$) spin in the $i$th unit cell, and $J$ and $J'$ are the effective couplings  derived from \cref{eq:effective-tj} by averaging over the groundstate charge distribution. 

The intra- ($J$) and inter- ($J'$) cell couplings are in general different.
However, in certain symmetrical cases, which we discuss in more detail further on, the couplings can indeed become identical.

In the generic case of $J\ne J'$, this periodic variation in the spin coupling opens a gap and gives rise to a dimerized state in the spin-sector in the strong-coupling limit \cite{schollwoeck2008quantum}. 
At the points $\delta = \pi /3, \pi, 5\pi/3$, described by the ionic $AB_2$ Hubbard model \cite{Torio2006}, the system has a site-centered inversion symmetry resulting in  $J=J'$ and the spin excitations again become gapless. 

To help illustrate the nature of these states, we consider two specific cases of $\delta$ in the atomic limit $U \gg V \gg t$.
For $\delta=0$ and in the atomic limit with $U \gg V \gg t$, 
the unit cell has two sites with energy $-3/4V$, coupled with inter-site tunneling $t$, and one site
with on-site potential $+3/4V$. 
At density $\rho =2/3$, the groundstate has both lower sites occupied and the energetically unfavorable site is unoccupied. 
This high-energy site can be adiabatically eliminated, resulting in an effective inter-cell tunneling $t' = 3t^2/2V$. 
The result is an intra-site spin coupling $J'=4t^2/U$, and a much lower inter-cell coupling $J=\frac{9t^4}{UV^2}$. 
Thus, here, the spin-sector is gapped.

For $\delta=\pi/3$, the potential structure of the unit-cell is inverted, compared to the $\delta=0$ case. 
This precise case has been studied in detail by~\citet{Torio2006}.
Here, we have one occupied site with onsite potential $-3V/4$, and two sites with energy $+3V/4$ sharing a fermion. 
This state has an inversion symmetry around the lower occupied site. 
Combined with the lattice-translation symmetry, this implies:
\begin{equation}
  J = J' \sim \frac{t^2U}{U^2 - \left(\frac{3V}{2}\right)^2}\,. 
\end{equation}
Thus, at points where $\delta \in \{1,3,5\}\pi/3$, the spin-dimerization disappears and the spin sector becomes gapless \cite{Aligia2004}.

The different phases in the strong- and weak-coupling limits can be understood in the context of symmetry-protected topological states. 
At the $\delta = \pi/3$ point, the model has a lattice-inversion symmetry around the first site in the unit cell. 
This lattice-inversion symmetry, combined with the $U(1)$ charge conservation, can give rise to a one-dimensional symmetry-protected topological phase. 
Such phases can be classified by the many-body Zak phase $\varphi$ [see Eq.~\eqref{eq:zak-phase}], which can only take  values of 0 and $\pi$ when the inversion symmetry exists.
We  consider the atomic limit, where $U,V \gg t$: In this limit, when $U \ll V$, we have the lowest site occupied with a spin singlet. 
This state has $\varphi = 0$. 
Alternatively, in the $U \gg V$ limit, the lowest site is occupied with one fermion and one inter-site dimer occupied. 
This phase has $\varphi = \pi$. 
As these states are characterized by different values of a  topological invariant (as long as the lattice-inversion symmetry is preserved), the many-body gap necessarily closes as the Hamiltonian is adiabatically transformed between the two limits.  

In summary, for $U\gg t, V$, the $\rho=2/3$ insulator has a gapped spin-dimerized groundstate, except for the special symmetric lines, where there is a gapless correlated-insulator phase. 

\section{Topological properties}
\label{sec:res_topo_props} 

\begin{figure}[tb]
    \centering
    \includegraphics[width=\columnwidth, clip]{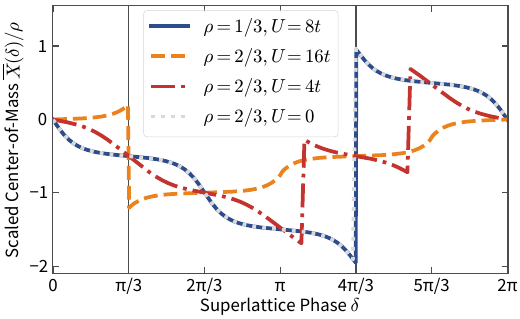}
    \mycaption{Center-of-mass motion.}{Scaled center-of-mass of an open system as a function of superlattice phase $\delta$. For  every cycle of $\delta\to\delta+2\pi$, there are discontinuities corresponding to minus the Chern number of the respective phase. The center-of-mass coordinate $\overline{X}$ [see \cref{eq:cm_motion}] is divided by particle density $\rho$ to show that the results for the Mott-insulating phase at $\rho=1/3$ agree with the behavior of the noninteracting phase for $\rho=2/3$. Data were obtained for length $L=60$. }
    \label{fig:center_of_mass}
\end{figure}

The family $\lbrace \hat H(\delta,\theta)\rbrace_{\mathbb{T}^2}$ of 1D models \cref{eq:fhh} inherits topological
properties from the 2D Harper-Hofstadter model for  $U=0$: At density
$\rho\in\{2,4\}/3$, the filled Hofstadter bands [see \cref{fig:bandstructure}] are topologically
nontrivial. 
Thus, a quantized amount of charge is pumped during a cycle $\delta \to
\delta + 2\pi$ in the {infinite} 1D model, \cref{eq:fhh}.

When we include interactions and the groundstate manifold remains gapped, we can compute Chern numbers as described in \cref{sec:fukui_method}. 
In this section we study topological properties of the interacting insulating phases discussed previously. 

As the Chern numbers are computed for a finite $L$ with periodic boundary conditions there are some subtle points we must address. 
Firstly, for finite systems there are no {gapless} excitations in any insulating phase. 
We can therefore compute Chern numbers from unique groundstates. 
However, spin excitation gaps in some of the Mott-insulating and the correlated-insulator phases close as $L\to\infty$. 
Therefore, the meaning of the Chern number in this limit, or equivalently, the stability of charge transport quantization in the related charge pump, is not guaranteed. 
We discuss this issue in more detail in \cref{sec:chern_gapless}.

\subsection{Mott insulator at density \texorpdfstring{$\rho=1/3$}{rho=1/3}}
\label{sec:top_mi_1_3}
 At density $\rho=1/3$, the lowest band is half-filled. 
 As discussed in \cref{sec:mi_1_3}, adding onsite interactions opens a charge gap for all $\delta,\,\theta$ and the phase becomes insulating. 
 While this phase has gapless spin excitations for $L\to \infty$, the groundstate manifold for finite systems is gapped already for $0<U\ll t$. 
Thus, we compute Chern numbers for this phase as described in \cref{sec:fukui_method}.

\begin{figure}[t]
    \centering
    \includegraphics[width=\columnwidth, clip]{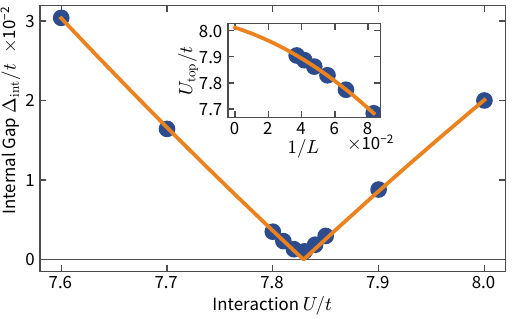}
    \mycaption{Internal gap for the topological transition.}{
        The Chern number of the groundstate manifold can only change when the energy gap closes.
        The density is $\rho=2/3$, results are for $\delta=\pi/3,\, \theta = 0$.
        Data are shown for length $L=18$, the line is a fit using the absolute value of a second-order polynomial.
        \emph{Inset:} {Critical} interaction strength $U_{\rm top}/t$ extrapolated to infinite system size using a quadratic fit in $1/L$ (orange line).}
    \label{fig:gap_closing}
\end{figure}

For all systems sizes $12 \leq L \leq 42$ considered here we find a Chern number $C(\rho=1/3, U>0)=-1=C_l$, 
which is equal to the Chern number $C_l$ of the lowest band
of the noninteracting Hofstadter model. 
We motivated this finding in \cref{sec:mi_1_3}:
{Single-particle} observables in the strongly-interacting phase approach the
expectation values for spinless, free fermions in the charge sector.
Therefore, we might expect to find the Chern number for a single species of free fermions, which would fill the lowest band, $C(\rho=1/3)=C_l$. 

We also compute the center-of-mass coordinate $\overline{X}$ in the strongly interacting regime as a function of $\delta$, shown in \cref{fig:center_of_mass}. 
We observe a single jump of the center-of-mass coordinate \cref{eq:cm_motion} by the negative value of the Chern number (solid line in the figure). 
The discontinuity is located at $\delta=4\pi/3$, 
when the two `lower' sites in the superlattice potential are energetically degenerate, see \cref{fig:realspace_model}.

Note that this center-of-mass curve perfectly lies on top of the one for the free model ($U=0$) at density $\rho=2/3$. 
This illustrates that the charge degrees of freedom in the strongly interacting phase behave much like a single component,  free Fermi gas, underlining our analogy with spinless fermions.

\subsection{Topological transition  at density \texorpdfstring{$\rho=2/3$}{rho=2/3}}

At density $\rho = 2/3$, we find that there are (at least) two different topological families, depending on the strength of the interaction $U$.
In \cref{sec:MI_2_3}, we saw that at density $\rho=2/3$ there are a number of phases in the $\{\delta,U\}$-parameter space. 
We find that the first of these phase transitions closely coincides with a topological transition in the Chern number. 
This transition occurs when the many-body gap closes in the $\{\delta,\theta\}$-manifold, at a critical $U = U_{\rm top}$, which in general is dependent on the system size, but should converge to $U_{c_1}$ in the $L\rightarrow \infty$ limit \cite{Torio2006}. 
Since the SDI phase is very narrow for our choice of parameters, we do not make any statement about the Chern number in the parameter regime that includes
the SDI phase.

\subsubsection{Weak interactions at density \texorpdfstring{$\rho=2/3$}{rho=2/3}}

Without interactions, the system is a band insulator that corresponds exactly to the groundstate of the Harper-Hofstadter model with the lowest band filled. 
Thus, the Chern number for density $\rho=2/3$ is given by two times the Chern number of the lowest Hofstadter band, $C(\rho=2/3)=-2$.
When we vary both interaction and superlattice potential strength, we consistently find a Chern number $C(\rho=2/3, U<U_{\rm top})=-2$. 

We find that for a finite lattice with periodic boundaries, the many-body gap closes only at one critical interaction strength $U_{\rm top}$, and only at three points on the $\{\delta, \theta \}$-manifold: $\delta\in\{1,3,5\}\pi/3,\,\theta= 0$, as shown for $\delta=\pi/3$ in \cref{fig:gap_closing}. 
In the thermodynamic limit, we expect this transition to occur when the system undergoes a phase transition to the SDI phase at $U_{c_1}$. 

The center-of-mass coordinate for the weakly-interacting system during the pump cycle is shown in \cref{fig:center_of_mass}
(dot-dashed line). 
The amplitude of the discontinuities agrees with the negative Chern number, i.e., $-C=2$.
The values of $\delta$ where the jumps occur are not directly related to the symmetry of the lattice,  but also depend on the interaction strength $U$.

\begin{figure}[tb]
    \centering
    \includegraphics[width=\columnwidth, clip]{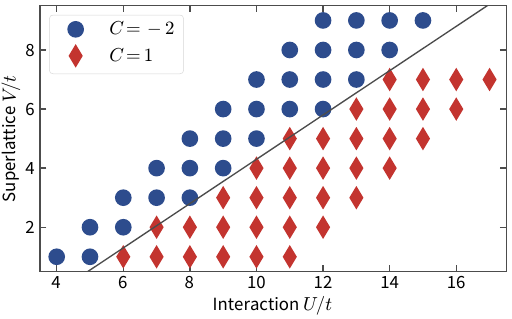}
    \mycaption{Topological transition for $\rho=2/3$.}{The Chern number changes from  $C(U<U_{\rm top})=-2\to C(U>U_{\rm top })=1$ for the insulators at $\rho=2/3$.
    Data are shown for length $L=18$ with twisted boundary conditions. The solid line is the transition $V=V(U_{\rm top})$.}
    \label{fig:top_transition}
\end{figure}

\subsubsection{Strong interactions at density \texorpdfstring{$\rho=2/3$}{rho=2/3} \label{sec:topo_strong_interaction}}
For strong interactions, with $U> U_{\rm top}$, the internal gap reopens for the entire groundstate manifold of $\{\hamil(\delta,\theta)\}_{\mathbb{T}^2}$ for finite system lengths (see \cref{fig:gap_closing}).
As shown in \cref{fig:top_transition}, we consistently find $C = +1$ in the presence of strong interactions. 
However, the global many-body gap closes at certain points due to the existence of gapless spin excitations.   

Our data indicate a linear dependence of $U_{\rm top}$ on the potential strength $V$.
For large $V/t\approx 30$, we find that the Chern number changes for $U\approx 3V/2$ (not shown in \cref{fig:top_transition}) which we would expect for the strong-coupling limit discussed in  \cref{sec:MI_2_3}.

Considering the center-of-mass coordinate for one cycle $\delta\to\delta+2\pi$ in an open chain, we find that strong interactions change sign and amplitude of the jumps in \cref{fig:center_of_mass}.
For strong interactions $U/t=16$ (dashed line in the figure), the discontinuity always occurs at $\delta=\pi/3$, the point of lattice-inversion symmetry, see \cref{sec:symmetries}.
Thus, the quantization of the pump cycle directly corresponds to the edge states of the SPT phase of the 1D chain at $\delta=\pi/3$.

Similar to the charge pump at density $\rho=1/3$ discussed in \cref{sec:top_mi_1_3}, we can  understand the change of the Chern number from band occupations. 
We found in \cref{sec:MI_2_3} that interactions suppress double occupation of lattice sites as well as double occupation of {bands}. 
Expectation values of single-particle observables in the limit $U\gg V \gg t$ thus approach the values for spinless fermions. 
Indeed, for spinless fermions at density $\rho=2/3$, we would expect the Chern number $C_l+C_m=-1+2=1$ which agrees with the numerically computed many-body Chern number.

\subsection{Interaction-induced degeneracies}
\label{sec:degeneracies}

\begin{figure}[tb]
    \centering
      \includegraphics[width=\columnwidth, clip]{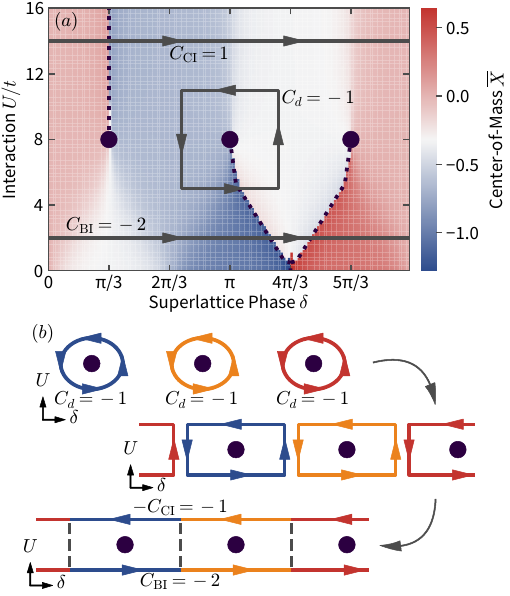}
      \mycaption{(Color online) Center-of-mass coordinate and topological structure for paths in the $\lbrace U,\delta\rbrace$-parameter space at $\rho=2/3$.}{ (a)  Quantized charge transport corresponds to discontinuities of $\overline{X}$ when computed from groundstates with open boundary conditions. 
      The purple circles at $\delta\in\{1,3,5\}\pi/3,\, U_{\rm top}/t=U/t\approx 8$  symbolize three topologically protected degeneracies in the case of $V/t=3$. 
    Any path encircling counter clockwise exactly one of these degeneracies has a Chern number $C_d=-1$, as it crosses exactly one jump changing $\overline{X}$ by $+1$
    (indicated by the dotted lines).  Data were obtained for length $L=60$ and open boundary conditions. 
(b) Three paths encircling one degeneracy each can be deformed and composed to form two separate paths; the $U<U_{\rm top}$ path and a $U>U_{\rm top}$ path. 
As this is a smooth deformation the sum of all Chern numbers cannot change, and the difference in Chern numbers between the two paths must be $3\times C_d=-3$. 
  } \label{fig:com_delta_U}
\end{figure}

The transition between the band and correlated-insulating states at density $\rho=2/3$ can be understood through certain interaction-induced degeneracies, as exemplified in  \cref{fig:gap_closing} for $\delta =\pi/3$. 
These degeneracies exist as points in the two-dimensional $\{U,\delta\}$-parameter space, which has the topology of a cylinder. 
The situation is illustrated in \cref{fig:com_delta_U}(a). 
The entire $\{U,\delta\}$-parameter space is simply connected through adiabatic transport. 
However, it is not possible to deform the entire closed path with $C_{\rm BI}=-2$ (in the band-insulating region) into the $C_{\rm CI}=1$ path (in the correlated region), as this would require crossing the degeneracies.

Each of these degeneracy  points has a Chern number $C_d=-1$ associated with it, corresponding to the path encircling the point in the $\{U,\delta\}$-parameter space,  as shown in \cref{fig:com_delta_U}(a). 
A pump cycle which encircles one of these points will transport a quantized charge of $-1$. 
This can be seen directly from the change in the center of mass value as one moves along this path  
in \cref{fig:com_delta_U}(a).

Finally, consider three paths each encircling one of these points in counter-clockwise direction, as shown in \cref{fig:com_delta_U}(b). 
These three paths can be composed to produce two paths, one for the band-insulating path, and one for the correlated-insulator path, but in the direction of $-\delta$.
For $C_{\rm BI} = -2$ the Chern number of the band insulator at this density, and $C_{\rm CI} = 1$ for the path in the large $U/t$ phase
that includes the CI phases.  
This implies that $C_{\rm BI} -C_{\rm CI} = 3C_d$, which is indeed the case.

\subsection{Chern numbers on gapless systems}
\label{sec:chern_gapless}

We must also address the question of the Chern number in the thermodynamic limit.
For finite systems, there are no {gapless} spin excitations in any insulating phase, such that the Chern numbers computed in the previous sections are well-defined.
However, in the case of Mott and correlated insulators, the spin excitations can become gapless as $L\to\infty$ (see the discussion in Secs.~\ref{sec:scl} and \ref{sec:strong_coupling2_3}).
This raises the question of the validity of such a topological classification in the thermodynamic limit: Do the gapless spin excitations preclude the possibility of adiabatic charge transport, or does the charge gap allow for quantized charge transport?

For the $\rho=1/3$ Mott insulator, the system does not pass through any phase boundaries. 
This would suggest that adiabatic charge transport is well-defined in this phase and remains quantized, reflecting the topology described in \cref{sec:top_mi_1_3}.

For the $\rho=2/3$ path in the strongly interacting regime, the system is gapped everywhere, except at the three points at $\delta=\pi/3,\pi,5\pi/3$ where there is a correlated insulator without spin-charge separation \cite{Torio2006}.
This state is also associated with a (weakly) divergent electric susceptibility \cite{Manmana2004,Tincani2009} which would suggest a possible breakdown of adiabatic transport when taking the system along this path.
As our present results do not provide further insight into these issues we leave them for future work.

Next, we  consider the consequence of these issues on the practical question of experimental observations.
It is expected that the topological properties of our model will manifest themselves in a quantized charge transport for ultracold atoms in an atomic lattice acting as a charge pump.
This has been recently demonstrated in the case of bosons \cite{Lohse2016} and fermions \cite{Nakajima2016}.
In both these cases, the experiment was conducted with spinless particles, in a completely gapped phase.
So far, there have been no such experiments with strongly interacting systems.
In these experiments, the charge transported was only approximately quantized, due to several factors:
finite-size effects, non-adiabaticity from finite pump time, technical heating and the presence of an harmonic trap.
As such, it is not clear that the fluctuations due to the spin degree of freedom would be discernible, particularly at very strong interaction strengths, where the prefactor of the electric susceptibility is expected to be very small \cite{Manmana2004}.
Moreover, quantum-gas experiments work with finite particle numbers of typically $N \sim 100$ atoms or less per one-dimensional system and charge pumps 
are performed only for a limited number of cycles \cite{Lohse2016,Schweizer2016}.
Therefore, we expect that an experiment would show the predicted transition at $\rho=2/3$ from $C=-2$ to $C=1$ during the accessible first pump cycles.

\section{Summary}
\label{sec:sum}

In this paper, we studied a one-dimensional fermionic lattice model with a superlattice potential and onsite repulsion.
For a  family of these systems defined on a torus of parameters, we can define a topological invariant which is invariant under small perturbations.
In the limit of large system sizes we can also interpret such families as topological Thouless charge pumps.
Without interactions,  the family of Hamiltonians maps directly  to the  2D Harper-Hofstadter model and thus is in the same topological class.

A particularly interesting situation arises at certain values of the superlattice phase, where, as a function of
$U/t$, a series of transitions exists, from a band insulator to a spontaneously dimerized insulator to a correlated insulator.
Theory and previous works \cite{Manmana2004,Torio2006,Tincani2009} predict that these transitions are Ising and BKT, respectively, which is consistent with our numerical data.
We argue that the first transition leads to a degeneracy in the full two-dimensional parameter space and a change of the Chern number
from $C=-2$  to $C=1$. The SDI phase is too narrow for the parameters considered here and hence we don't make a statement about the
Chern number there. This change of the Chern number can be understood from simple intuitive arguments in the atomic limit resulting from
a competition of the superlattice potential strength $V$ with the interaction strength $U$.
The change of the Chern number is clearly seen in our finite-size data and we expect that this $U$-driven transition should
be detectable in a charge-pumping experiment. Different from the fermionic Rice-Mele model \cite{Nakagawa2018}, we don't observe a breakdown
of the charge pump when studying the same quantities as in \cite{Nakagawa2018} on finite system sizes.
The presence of gapless spin excitations along special points of the  pump cycle parameterized by $\delta$ may ultimately spoil the quantization of $C$ at large $U/t$, but we expect that for the first pump cycles that can typically be accessed in a quantum-gas experiment,
$C$, and hence the pumped charge, should remain quantized. The clarification of this question, theoretically related to the degree of spin-charge separation, and its investigation in time-dependent simulations is left for future research.

There might be challenges for the experimental realization in the regimes with correlated insulators  due to the gapless spin excitations.
The vanishing or small finite-size many-body gaps may pose constraints on the pump speed.
Further research is necessary to determine the optimum timescales for adiabatic pumping in strongly interacting phases
in time-dependent simulations.

We thank V. Gurarie, S. Manmana, and R. Verresen  for useful discussions. This work was supported by the Deutsche Forschungsgemeinschaft (DFG, German Research
Foundation) via DFG
Research Unit FOR 2414 under project number 277974659 and under Germany's Excellence Strategy -- EXC-2111 -- 390814868.
Part of this research was conducted at KITP at UCSB.
This research was supported in part by the National Science Foundation under Grant No.~NSF PHY-1748958.
This work was further supported by ERC Grant QUENOCOBA, ERC-2016-ADG (Grant no. 742102).

\bibliographystyle{apsrev4-1}
\bibliography{fhh_paper}

\appendix

\begin{figure}[h!]
    \centering
    \includegraphics[width=\columnwidth, clip]{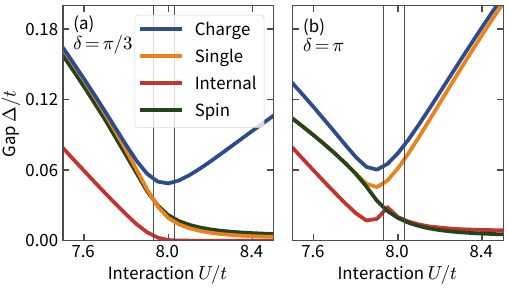}
    \mycaption{Gaps at the phase transitions and inversion symmetry.}
    {Data were computed for open boundaries at $L=600$ and $\rho=2/3$.
    We consider (a) $\delta=\pi/3$, which is inversion {symmetric} and (b) $\delta=\pi$ which is not inversion symmetric.
    Edge states only exist for $\delta=\pi/3$, explaining the qualitatively different behavior of the single-particle gap in the large-$U$ limit.
     }
    \label{fig:4gaps_boundaries}
\end{figure}

\section{Bulk and edge symmetry \label{sec:edge-symm-appendix}}

In this appendix, we provide more details regarding the edge effects for our open-boundary condition data. 
As is well-known from DMRG studies \cite{Qin1995}, the particular choice of lattice termination can have important effects on the excitation spectrum. 
This is directly related to the presence of gapless edge states in symmetry-protected topological states \cite{Essin2011}. 
In our model, we have this situation in the $\rho=2/3$ insulating phases, where the system has additional lattice symmetries along $\delta=2\pi/6 + 2\pi n/3$. 

When considering the three-site superlattice with open boundaries, a shift of the superlattice phase $\delta$ by $2\pi/3$ changes the properties of the edge:
For the configuration sketched in \cref{fig:realspace_model}, a shift $\delta\to\delta+2\pi/3$ removes the lattice-inversion symmetry, such that we have two energetically higher sites on one end.
This explains why the discontinuities in \cref{fig:center_of_mass}, related to edge states, do not have the same symmetry as the bulk.
Introducing the single-particle gap,
\begin{multline}
    \Delta_{\rm single}(N) = E(N+1, S=1/2)  \\+ E(N-1, S=1/2) - 2 E(N, S=0)\,,
\end{multline}
we observe in \cref{fig:4gaps_boundaries} that the degenerate edge states only appear for a `symmetric' choice of boundaries.
We observe in the same plot how the spontaneously dimerized phase (SDI) is prohibited by asymmetric boundaries: 
Only one dimerization is allowed and thus the internal energy gap has a local maximum for $U/t\approx 7.9$, when long-range dimer order appears.

\begin{figure}[t!]
    \centering
    \includegraphics[width=\columnwidth, clip]{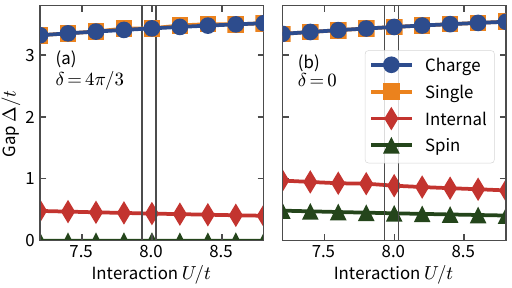}
    \mycaption{Gaps at other values of $\delta$.}
    {Data were computed for open boundaries at $L=600$ and $\rho=2/3$, corresponding to \cref{fig:4gaps_boundaries}, however, the superlattice is shifted via $\delta\to\delta+\pi$ corresponding to $V\to -V$.
    We consider (a) $\delta=4\pi/3$, which is inversion symmetric and (b) $\delta=0$ which is not inversion symmetric.
    The inversion-symmetric lattice hosts a gapless spin mode at the boundaries, leading to the vanishing spin gap.
    }
    \label{fig:4gaps_boundaries_acrit}
\end{figure}

\begin{figure}[t!]
    \centering
    \includegraphics[width=\columnwidth, clip]{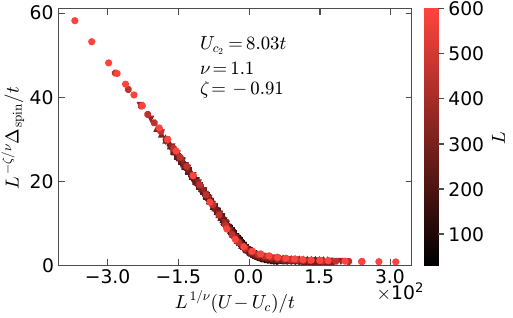}
    \mycaption{Scaling collapse for the spin gap at $\delta=\pi/3$.}{Data and colors are the same as for \cref{fig:gaps_scaled}(b) and  the parameters $\nu,\zeta,U_{c_2}$ for the finite-size scaling collapse were determined numerically from Eq.~\eqref{eq:scaling}.
  The system sizes used in the scaling collapse are $L=30,60,90,120,150,180,210,240,270,300,360,450,600$.  }
    \label{fig:spin_collapse}
\end{figure}

In Fig.~\ref{fig:4gaps_boundaries_acrit}, we show the four gaps (charge, single-particle, spin and internal gap) as a function of $U/t$
for $\delta=0$ and $4\pi/3$. At these values, there is no gap closing as $U/t$ increases (compare Fig.~\ref{fig:rho23-phase-diagram}).
We observe that the charge and single-particle gap are identical for the values of $U/t$ considered in the figure. The spin gap
is the smallest gap in both cases, while the internal gap exhibits a weak decrease with $U/t$. For $\delta=4\pi/3$, which is inversion
symmetric, there exist spin-edge modes and therefore, the spin gap vanishes.
For systems with open boundary conditions, the band insulator has gapless edge states at $\delta = 4\pi/3$.
The location of these edge states smoothly changes with increasing $U$.

Finally, Fig.~\ref{fig:spin_collapse} shows a scaling collapse of finite-size data for the spin gap at $\delta=\pi/3$ using Eq.~\eqref{eq:scaling}.
This results in an estimate of the critical $U_{c_2}/t\approx 8.03$ of the second transition. This transition is predicted to be
of BKT type \cite{Manmana2004, Torio2006} [as supported by the data shown in Fig.~\ref{fig:gaps_scaled}(b)], yet our system sizes are not large enough to reliably extract the
critical value from a BKT scaling and a conventional scaling analysis works as well. Therefore, $U_{c_2}/t\approx 8.03$ has to be understood as a lower bound to the actual critical value.

\clearpage

\end{document}